\documentclass[10pt]{article}
\usepackage{amsmath,amssymb,amsfonts,amsthm,bm}
\usepackage[margin=1in]{geometry}
\usepackage{graphicx}
\usepackage[pdftex,colorlinks=true,linkcolor=black, citecolor=black, urlcolor=black,letterpaper,bookmarks=true,bookmarksopen=false,bookmarksnumbered=false,hypertexnames=true]{hyperref}
\usepackage[round,comma]{natbib}

\newcommand{\mat}[1]{\text{\bf #1}}

\begin{document}
\title{A note on the formulation of the Ensemble Adjustment Kalman Filter}
\author{Ian Grooms}
\date{Department of Applied Mathematics, University of Colorado, Boulder, Colorado, USA 80309}
\maketitle

\begin{abstract}
The ensemble adjustment Kalman filter (EAKF; Anderson, 2001) is one of the earliest ensemble square root filters.
This note clarifies the correct formulation of the EAKF, which depends on a careful treatment of an eigen-decomposition of one of the matrices involved in the formulation.
\end{abstract}

\vspace{\baselineskip}
Ensemble Kalman filters (EnKFs) are widely used for high-dimensional, nonlinear, non-Gaussian data assimilation.
EnKFs use an ensemble to approximate the forecast mean and covariance, and then update these in a manner consistent with the classical Kalman Filter formulas \citep{Evensen09}.
The ensemble adjustment Kalman filter \citep[EAKF;][]{Anderson01} is one of the earliest EnKFs.
The EAKF provides a deterministic update of the forecast ensemble, unlike the perturbed-observation version of the EnKF \citep{BvLE98,HM98}; as such it is a member of the class of ensemble square root filters \citep{TABHW03}.
The presentation of the EAKF in \citet{Anderson01} and \cite{TABHW03} suffers from a lack of clarity potentially leaving the practitioner confused as to the correct implementation of the EAKF as well as confused about whether the EAKF is in fact consistent with the Kalman filter update formula.
This note clarifies the correct formulation of the EAKF, and presents a novel detail required to ensure consistency of the update formula with the Kalman filter covariance update.
No clarification is needed in the serial assimilation algorithm of \citet{Anderson03}, which bypasses the matrix algebra of the EAKF.\\

We follow the notation from \citet{TABHW03}.
Denote the forecast ensemble by $\bm{x}_1^f,\cdots,\bm{x}_m^f$ where $m$ is the ensemble size.
The ensemble forecast covariance matrix is $\mat{P}^f = \mat{Z}^f\mat{Z}^{fT}$ where $\mat{Z}^f$ is the scaled ensemble pertubation matrix
\begin{equation}
\mat{Z}^f = \frac{1}{\sqrt{m-1}}\left[\bm{x}_1^f-\bm{\mu}^f,\cdots,\bm{x}_m^f-\bm{\mu}^f\right]
\end{equation}
and $\bm{\mu}^f$ is the ensemble forecast mean.
Denote the observation matrix by $\mat{H}$ and the observation error covariance matrix by $\mat{R}$.

The key to the EAKF is the formula for updating the ensemble perturbations in such a way that they are consistent with the KF covariance update formula
\begin{equation}\label{eqn:KF}
\mat{P}^a=\left(\mat{I}-\mat{KH}\right)\mat{P}^f
\end{equation}
where $\mat{K}$ is the Kalman gain matrix
\begin{equation}
\mat{K} = \mat{P}^f\mat{H}^T\left(\mat{HP}^f\mat{H}^T+\mat{R}\right)^{-1}.
\end{equation}
Defining $\mat{V} = \left(\mat{HZ}^f\right)^T$ and inserting $\mat{P}^f = \mat{Z}^f\mat{Z}^{fT}$, the KF update formula becomes
\begin{equation}
\mat{P}^a=\mat{Z}^f\left[\mat{I}-\mat{V}\left(\mat{V}^T\mat{V}+\mat{R}\right)^{-1}\mat{V}^T\right]\mat{Z}^{fT}.
\end{equation}
The Woodbury matrix identity can be applied to write this as
\begin{equation}\label{eqn:EAKF}
\mat{P}^a=\mat{Z}^f\left[\mat{I}+\mat{V}\mat{R}^{-1}\mat{V}^T\right]^{-1}\mat{Z}^{fT}.
\end{equation}
The EAKF uses an adjustment matrix $\mat{A}$ to update the scaled ensemble perturbation matrix as $\mat{Z}^a = \mat{AZ}^f$ in such a way that $\mat{P}^a = \mat{Z}^a\mat{Z}^{aT}$ exactly satisfies Eqn.~(\ref{eqn:EAKF}).
The EAKF adjustment matrix presented by \citet{TABHW03} is
\begin{equation}\label{eqn:A}
\mat{A} = \mat{Z}^f\mat{C}\left(\mat{I}+\mathbf{\Gamma}\right)^{-1/2}\mat{G}^{-1}\mat{F}^T
\end{equation}
where $\mat{Z}^f = \mat{FGU}^T$ is the singular value decomposition (SVD) of $\mat{Z}^f$ and $\mat{V}\mat{R}^{-1}\mat{V}^T = \mat{C}\mathbf{\Gamma}\mat{C}^T$ is the eigenvalue decomposition of $\mat{V}\mat{R}^{-1}\mat{V}^T$ (cf.~Eq.~(19) of \citet{TABHW03}).

The use of the notation $\mat{G}^{-1}$ in (\ref{eqn:A}) implies that the expression uses the version of the SVD that produces a square, invertible matrix $\mat{G}$.
If the state dimension is $n$ and the rank of $\mat{Z}^f$ is $r$ then $\mat{Z}^f$ is $n\times m$, and the SVD produces $\mat{G}$ of size $r\times r$, $\mat{F}$ of size $n\times r$, and $\mat{U}$ of size $m\times r$.
This is precisely the interpretation given in a footnote by \citet{TABHW03}.
In this interpretation the only possible way for (\ref{eqn:A}) to be meaningful is for $r=m$, because otherwise the matrix $\left(\mat{I}+\mathbf{\Gamma}\right)^{-1/2}$, which is $m\times m$, can't be multiplied by $\mat{G}^{-1}$, which is $r\times r$.
But the rank of $\mat{Z}^f$ is $r\le$ min$\{n,m-1\}$, so (\ref{eqn:A}) is a mathematical impossibility, and the practitioner is left uncertain the correct formulation of the EAKF adjustment matrix.

The situation can be fixed by using the version of the SVD that produces  $\mat{G}$ of size $r\times m$, $\mat{F}$ of size $n\times r$, and $\mat{U}$ of size $m\times m$, and by replacing $\mat{G}^{-1}$ in (\ref{eqn:A}) by the Moore-Penrose pseudoinverse $\mat{G}^\dag$.
Here $\mat{G}^\dag$ is an $m\times r$ diagonal matrix whose diagonal elements are $1/\sigma_1,\ldots,1/\sigma_r$ where $\sigma_i$ are the singular values of $\mat{Z}^f$ that appear on the diagonal of $\mat{G}$.
With this substitution the ensemble adjustment matrix becomes
\begin{align}\label{eqn:Adag}
\mat{A} &=\mat{Z}^f\mat{C}\left(\mat{I}+\mathbf{\Gamma}\right)^{-1/2}\mat{G}^{\dag}\mat{F}^T.
\end{align}
This formulation is implied by the discussion of rank deficiency in appendix A of \citet{Anderson01}, though the consistency of this adjustment matrix with the KF covariance update formula is not fully demonstrated when $\mat{G}$ is not invertible.

We next show that this update formula leads to an analysis ensemble covariance matrix that exactly satisfies (\ref{eqn:KF}), as long as the matrices $\mat{C}$ and $\mathbf{\Gamma}$ are constructed carefully.
We begin by simplifying the expression for the scaled analysis ensemble perturbation matrix
\begin{align}\notag
\mat{Z}^a&=\mat{Z}^f\mat{C}\left(\mat{I}+\mathbf{\Gamma}\right)^{-1/2}\mat{G}^{\dag}\mat{F}^T\mat{FGU}^T\\
&= \mat{Z}^f\mat{C}\left(\mat{I}+\mathbf{\Gamma}\right)^{-1/2}\mat{G}^\dag\mat{G}\mat{U}^T.
\end{align}
The matrix $\mat{G}^\dag\mat{G}$ in this expression is an $m\times m$ diagonal matrix whose first $r$ diagonal entries are 1, and whose remaining entries are all 0.
The product $\left(\mat{I}+\mathbf{\Gamma}\right)^{-1/2}\mat{G}^\dag\mat{G}$ is a diagonal matrix whose first $r$ entries are the same as those of $\left(\mat{I}+\mathbf{\Gamma}\right)^{-1/2}$, and whose remaining entries are all 0.
Denoting this product by $\left(\mat{I}+\mathbf{\Gamma}\right)_r^{-1/2}$ we plug in to the update formula to find
\begin{equation}
\mat{P}^a=\mat{Z}^a\mat{Z}^{aT} = \mat{Z}^f\mat{C}\left(\mat{I}+\mathbf{\Gamma}\right)_r^{-1}\mat{C}^T\mat{Z}^{fT}.
\end{equation}
The correct expression differs only in the $_r$ subscript, meaning that in this formulation of EAKF the last $m-r$ elements on the diagonal of $\left(\mat{I}+\mathbf{\Gamma}\right)_r^{-1}$ are zero when they should not be.
This suggests that the EAKF produces a posterior covariance that is under-dispersed compared to the correct posterior covariance.
Nevertheless, this version of the EAKF still does produce the correct posterior covariance if implemented carefully, as we now show.

First note that $\mat{VR}^{-1}\mat{V}^T = \mat{Z}^{fT}\mat{H}^T\mat{R}^{-1}\mat{HZ}^f$, so the null space of $\mat{Z}^f$ is equal to the null space of $\mat{VR}^{-1}\mat{V}^T$.
This implies that there are $m-r$ columns of $\mat{C}$ (the columns of $\mat{C}$ are orthonormal eigenvectors of $\mat{VR}^{-1}\mat{V}^T$) that are in the null space of $\mat{Z}^f$.
Multiplying any one of these columns by $\mat{Z}^f$ will yield $\bm{0}$, so the matrix $\mat{Z}^f\mat{C}$ that appears in (\ref{eqn:Adag}) will have $m-r$ columns equal to $\bm{0}$.
If the eigenvalue decomposition of $\mat{VR}^{-1}\mat{V}^T$ is constructed so that these columns of $\mat{C}$ are the last $m-r$ columns of $\mat{C}$, then we have
\begin{equation}
\mat{Z}^f\mat{C}\left(\mat{I}+\mathbf{\Gamma}\right)_r^{-1/2}=\mat{Z}^f\mat{C}\left(\mat{I}+\mathbf{\Gamma}\right)^{-1/2}
\end{equation}
because the $m-r$ trailing zeros on the diagonal of $\left(\mat{I}+\mathbf{\Gamma}\right)_r^{-1/2}$ end up multiplying columns of $\mat{Z}^f\mat{C}$ that are already $\bm{0}$.
If, on the other hand, the $m-r$ eigenvectors in the null space of $\mat{VR}^{-1}\mat{V}^T$ are not arranged as the last $m-r$ columns of $\mat{C}$, then the $m-r$ trailing zeros on the diagonal of $\left(\mat{I}+\mathbf{\Gamma}\right)_r^{-1/2}$ end up multiplying nonzero columns of $\mat{Z}^f\mat{C}$, causing erroneous cancellations and leading to an under-dispersed posterior ensemble.
If the columns of $\mat{C}$ are arranged correctly, then we have
\begin{align}\notag
\mat{P}^a&=\mat{Z}^a\mat{Z}^{aT} = \mat{Z}^t\mat{C}\left(\mat{I}+\mathbf{\Gamma}\right)_r^{-1}\mat{C}^T\mat{Z}^{fT}\\\notag
&= \mat{Z}^t\mat{C}\left(\mat{I}+\mathbf{\Gamma}\right)^{-1}\mat{C}^T\mat{Z}^{fT}\\
&=\mat{Z}^f\left[\mat{I}+\mat{V}\mat{R}^{-1}\mat{V}^T\right]^{-1}\mat{Z}^{fT}
\end{align}
i.e. the EAKF update exactly satisfies (\ref{eqn:KF}).

The presentation of the EAKF in appendix A of \citet{Anderson01} avoids this potential problem by using the language `singular value decomposition' for the eigenvalue decomposition of $\mat{V}\mat{R}^{-1}\mat{V}^T$. 
Since $\mat{V}\mat{R}^{-1}\mat{V}^T$ is a symmetric non-negative definite matrix, the SVD is the same as the eigenvalue decomposition.
The crucial difference is that the convention for the SVD is to always arrange vectors corresponding to the null space as the last columns of the eigenvector matrix $\mat{C}$.
Eigenvalue solvers often do not follow this convention, and scatter the null space eigenvectors randomly in the columns of $\mat{C}$.

In summary, the correct EAKF update formula is
\begin{equation}
\mat{Z}^a = \mat{A}\mat{Z}^f,\;\mat{A} = \mat{Z}^f\mat{C}\left(\mat{I}+\mathbf{\Gamma}\right)^{-1/2}\mat{G}^{\dag}\mat{F}^T
\end{equation}
where $\mat{Z}^f = \mat{FGU}^T$ is the singular value decomposition (SVD) of $\mat{Z}^f$ and $\mat{V}\mat{R}^{-1}\mat{V}^T = \mat{C}\mathbf{\Gamma}\mat{C}^T$ is the eigenvalue decomposition of $\mat{V}\mat{R}^{-1}\mat{V}^T$, with the specific requirement that eigenvectors of $\mat{V}\mat{R}^{-1}\mat{V}^T$ that are in the null space of $\mat{V}\mat{R}^{-1}\mat{V}^T$ must be arranged as the final columns of $\mat{C}$.
The purpose of this note is to (i) fully demonstrate the consistency of the EAKF of \citet{Anderson01}, (ii) clarify the correct treatment of the pseudoinverse in the discussion of \citet{TABHW03}, and (iii) point out a potential pitfall in the implementation of the EAKF that could lead to inconsistency and an under-dispersed analysis ensemble if an eigenvalue decomposition is not treated carefully.


\end{document}